\documentclass{iopart}
\usepackage[latin9]{inputenc}
\usepackage{textcomp}
\usepackage{amstext}
\usepackage{amssymb}
\usepackage{graphicx}

\makeatletter

\newcommand{\noun}[1]{\textsc{#1}}
\newcommand{\lyxmathsym}[1]{\ifmmode\begingroup\def\b@ld{bold}
  \text{\ifx\math@version\b@ld\bfseries\fi#1}\endgroup\else#1\fi}

\providecommand{\tabularnewline}{\\}

\usepackage{iopams}
\usepackage{setstack}

\usepackage{dcolumn}% Align table columns on decimal point
\usepackage{bm}% bold math
\usepackage{hyphenat}\usepackage{enumerate}\usepackage{breqn}\usepackage{float}%\usepackage{dmath}

\makeatother

\begin{document}

\title{Prisoner's dilemma on directed networks}

\author{A. L. Ferreira}

\address{Departamento de F\'{i}sica$,$ I3N$,$ Universidade de Aveiro$,$
3810-193 Aveiro$,$ Portugal.}

\ead{alf@ua.pt}

\author{A. Lipowski}

\address{Faculty of Physics$,$ Adam Mickiewicz University$,$ Pozna\'{n}$,$
Poland.}

\ead{lipowski@amu.edu.pl}

\author{T. B. Pedro}

\address{Departamento de F\'{i}sica$,$ Universidade Federal de Santa Catarina$,$
Florian\'opolis$,$ 88040-900 Santa Catarina$,$ Brazil. }

\ead{tiago.b.p@posgrad.ufsc.br}

\author{M. Santos}

\address{Departamento de F\'{i}sica$,$ Universidade Federal de Santa Catarina$,$
Florian\'opolis$,$ 88040-900 Santa Catarina$,$ Brazil. }

\ead{marcio.santos@ufsc.br}

\author{W. Figueiredo}

\address{Departamento de F\'{i}sica$,$ Universidade Federal de Santa Catarina$,$
Florian\'opolis$,$ 88040-900 Santa Catarina$,$ Brazil. }

\ead{wagner.figueiredo@ufsc.br}
\begin{abstract}
We study the prisoner's dilemma model with a noisy imitation evolutionary
dynamics on directed out-homogeneous and uncorrelated directed random
networks. An heterogeneous pair mean-field approximation is presented
showing good agreement with Monte Carlo simulations in the limit of
weak selection (high noise) where we obtain analytical predictions
for the critical temptations. We discuss the phase diagram as a function
of temptation, intensity of noise and coordination number of the networks
and we consider both the model with and without self-interaction.
We compare our results with available results for non-directed lattices
and networks.
\end{abstract}
\maketitle

\section{Introduction}

Dilemmas arise when it is advantageous for an individual to act selfishly
by not taking into account the overall performance of the group it
belongs to\cite{Kollock1998,nowak2006,Szabo2007,sigmund2010}. Such
dilemmas can be modeled by simple two agent games where each agent
follows two strategies, cooperation (C) and defection (D) and receive
a payoff after each round that depends on the chosen strategy. In
the prisoners dilemma game\cite{Axelrod1981,Axelrod1984} (PD) a defector
facing a cooperator receives the highest payoff (T- temptation) higher
than the second best payoff (R- reward) received from mutual cooperation.
The smallest payoff is received by a cooperator facing a defector
(S - sucker payoff), even smaller than the payoff received from mutual
defection (P- punishment). In a population of defectors, an agent
does not increase it's payoff by changing it's strategy to cooperation
contrary to the case of a population of cooperators where it is advantageous
for an agent to change to defection. Defection is the best reply to
defection which classifies this strategy as a Nash equilibrium in
a game theory framework\cite{nowak2006}. However, cooperation is
observed in social and biological systems even when from an agent
point of view it is better not to cooperate. Evolutionary games are
obtained when the best strategies become more frequent by giving a
higher reproductive ability to agents following those strategies.
Several mechanisms for the evolution of cooperation were proposed,
some of them relying on strategies that assume repeated encounters
and some sort of memory \cite{nowak2006b}. 

Considering only pure strategies (C and D), it was found that the
introduction of a spatial or a network structure in the population
\cite{nowak1992,nowak1993,nowak1994,Lindgren1994,Hauert2005} where
an agent plays with nearest neighbors (possibly including itself)
and receives a  total payoff collected from each of those two agent
games can lead to the emergence of cooperation. 

The PD game has been studied in structured populations such as lattices\cite{nowak1992,nowak1993,nowak1994,Szabo1998,Szabo2005},
hierarchical lattices\cite{Vukov2005}, empirical social networks\cite{Holme2003},
small world networks\cite{Abramson2002,Kim2002,masuda2003,Tomochi2004,Santos2005,pacheco2005,Tang2006},
homogeneous random networks\cite{Vukov2006}, single scale \cite{Santos2006,Tang2006}
and scale-free networks\cite{pacheco2005,Santos2005b,Santos2006,Santos2006b,Tang2006}.
In the case of well mixed populations and fully connected networks
where the payoff of each agent depends directly on the frequency of
each strategy in the whole population the PD model evolves to a phase
where all agents are defectors. The clustering of cooperators in a
spatial/network structure allows cooperators to resist exploitation
and the population may evolve to a coexistence phase of cooperators
and defectors\cite{Ohtsuki2006}. Structural heterogeneity in the
number of neighbors was found to favor generically cooperation in
the PD game \cite{Santos2005,nowak2006b,Santos2006,Santos2006b} although
this effect may depend on the details of the adopted evolutionary
rule \cite{Szabo2007,maciejewski2014}. 

The distinction between the lattice/network of interactions where
the agents play and the lattice/network of influence where the learning
and reproduction process takes place\cite{Ohtsuki2007,Ohtsuki2007b,Wu2007}
was previously introduced. In those studies both the network of interactions
and the network of influences was considered to be non-directed. The
influence between agents in empirical social networks is sometimes
found to be asymmetric and requiring modeling by directed networks\cite{Guha_2004}.
This asymmetry may be taken into account in the framework of evolving
networks by considering time evolving weights which depend on the
outcome of ongoing interactions between the agents\cite{Wu2006,Szabo2007,Skyrms2000,Lipowski2014b}
.

In this work we study the PD model in the case where the network of
interactions and the network of learning are identical directed networks.
Specifically, we consider that when an agent A plays with it's neighbor
agent B ( which does not have A as its own neighbor) only A collects
the payoff for the game between them. Even though the PD game is a
two agent game the payoffs (consequences) of the adopted strategies
(actions) may be collected asymmetrically by the two players engaged
in a particular interaction. In previous works the case of a single
influential node with long-range asymmetric interactions was considered\cite{Kim2002}
and the evolutionary dynamics on a directed cycle (one dimensional
directed lattice) where the payoffs of the two agent games are collected
asymmetrically was discussed\cite{Lieberman2005}.

Several techniques were used to study two player game models on spatial
structures such as Monte Carlo simulation, calculations of the probability
of fixation of mutant strategies\cite{Ohtsuki2006} and the determination
of the evolutionary stable strategies by deriving an effective replicator
dynamics\cite{Ohtsuki2006b,Ohtsuki2008}. The last two techniques
assume the limit of weak selection and rely on mean-field (MF) pair
approximation. \cite{Szabo1998,Szabo2007}. Unless in conditions of
weak selection, higher order cluster MF approximations\cite{Szabo1998,Vukov2006,Szabo2007},
going beyond the pair approximation, are usually needed to obtain
accurate results. Such higher order approximations were applied before
to model dynamical processes in regular lattices\cite{Ferreira1993,Avraham1992,Szabo1998,Szabo2007}
and networks with an homogeneous degree structure\cite{petermann2004}.
In directed lattices with a local tree like structure single-site
MF approximations may give surprisingly accurate results\cite{Lipowski2015}.
In this work, we present an heterogeneous single-site and pair MF
approximation for the PD game on a generic directed network taking
into account the degree heterogeneities and degree correlations. In
general, there is no unique way of deriving mean-field approximations
from the master equation describing dynamical processes in the networks\cite{Dorogovtsev2008,Vespignani2012}.
Our approach is close in spirit to the annealed mean-field approximations
previously used to study epidemic spreading in heterogeneous networks\cite{pastorsatorras2001,mata2014}. 

The remaining of the paper is organized as follows: in section \ref{2}
we describe in detail the version of the PD model and the networks
considered; in section \ref{3} we present our single-site and pair
MF approximations; in section \ref{4} the results obtained for out-homogeneous
directed networks of different out-degree are presented; in section
\ref{sec:5} we compare the predictions of the pair approximation
with Monte Carlo simulation for the steady-state density of cooperators
in Poissonian random directed networks and finally in section \ref{sec:6}
we summarize our main conclusions.

\section{the model on directed random networks}

\label{2}

We consider a scaled version of the payoff matrix of the PD model
\cite{nowak1992,Szabo1998} where the reward payoff for mutual cooperation
is set to unity, the temptation payoff received by a $D$ facing a
$C$ is $b$ and all other payoffs are null. The dilemma strictly
exists for $b>1$ when defecting becomes advantageous from the individual
point of view. When the game is played on a network, at each vertex,
$i$ , there is an agent that receives a total payoff, $P_{i}$, that
depends on the strategies adopted by the neighbors. We study in more
detail the case where self-interaction is included and the payoff
for a $C$ is then equal to $n_{C}+1$ and the payoff for a $D$ is
equal to $b\, n_{C}$, being $n_{C}$ the number of neighbors that
follow the strategy $C$. The inclusion of self-interaction may be
justified by seeing each agent as representing a group following a
given strategy\cite{nowak1992} in a coarse-grained sense. Unless
explicitly mentioned the results presented are for the model with
self-interaction. For a network with $N$ vertices, where a vertex
$i$ has $k_{i}$ neighbors, an \emph{imitation dynamics} is considered
such that a randomly chosen agent, $i,\,1\leq i\leq N$ chooses a
random neighbor, $i_{\mu},1\leq\mu\leq k_{i}$, and if this neighbor
is following a different strategy it adopts the strategy of the neighbor
with probability 

\begin{equation}
p_{\mathrm{imitate}}(P_{i},P_{i_{\mu}})=\frac{1}{1+\exp\left(-\beta\left(P_{i_{\mu}}-P_{i}\right)\right)},\label{eq: p_imit}
\end{equation}
where $T=\beta^{-1}$ is a temperature like parameter that controls
the level of noise in the strategy imitation process. The weak selection
limit corresponds to large $T$ when the dependence of the imitation
probability on the agent's payoff difference is small and linear.
The above imitation dynamics\cite{Szabo1998} is sometimes called
pairwise comparison\noun{ }dynamics and it is just one of several
possible reasonable evolutionary dynamics \cite{Ohtsuki2006c}. In
a system where each agent has $z$ neighbors the total payoff of the
full cooperation phase would be $N\left(z+1\right)$ and an alternating
phase where each $C$ is surrounded by $D$ and vice-versa, if possible,
would have a total system payoff $Nbz/2+N/2$. Thus only for $b<2+1/z$
($b<2$ without self-interaction) the full cooperation phase corresponds
to the maximum system payoff but, for any value of $b$, the full
defection phase is always the phase with the smallest system payoff.
In a large fully connected network the density of cooperators $\rho(t)$
can be written as $d\rho/dt=\rho\left(1-\rho\right)\tanh\left(\beta\rho(1-b)\right)$
which, in the limit of weak selection, has the replicator equation
form \cite{nowak2006,sigmund2010} with a solution that reaches zero
for long times, when $b>1$.

We study the model on two kinds of directed random networks where
the neighbors of a vertex are connected through outgoing links starting
from a vertex. In the directed out-homogeneous networks\cite{Lipowski2014,Lipowski2015}
the number of out-links, $z$, is the same for all vertices. The out-links
are generated by selecting randomly, for each vertex of the network,
$z$ other (different) vertices. The distribution of in-links is Poissonian
as for random networks. For $z>1$ the system is in a percolating
phase with a giant strongly connected component containing a finite
fraction of the vertices\cite{Lipowski2015}. The other kind of networks
considered are directed random networks which are built by establishing
an outgoing link, $i\rightarrow j$ , from each vertex, $i$, to each
other vertex, $j$ with probability $q/N$. The distribution of the
number of outgoing links of a given vertex as well as the distribution
of the number of ingoing links is Poissonian with an average value
$q.$ For these networks, vertices with no out-links (without neighbors)
may receive an arbitrary number of in-links, thus influencing other
vertices, while their own strategies do not suffer influence from
others. To avoid this behavior we generated the out-links starting
from a truncated and renormalized Poisson distribution where vertices
with no out-going links have zero probability. The average number
of out-links of a vertex is then $z=\left\langle k_{\mathrm{out}}\right\rangle =\tfrac{q}{1-\exp(-q)}$,
always larger than unity. To generate these networks we use a method
based on the configuration model\cite{molloy1995}: the out-degrees
of each vertex, $i$, are drawn from its probability distribution,
thus generating $k_{i}$ stubs which are ends of outgoing links emerging
from the vertex $i$. These stubs are then connected to randomly chosen
vertices with the restriction of not repeating a vertex and not allowing
connections to itself. 

The directed networks\cite{Dorogovtsev2001,boguna2005,Serrano2009}
are characterized by a joint in-degree and out-degre distribution,
$P\left(k_{\mathrm{i\mathrm{n}}},k_{\mathrm{out}}\right)$ and degree
correlations $P_{\mathrm{i\mathrm{n}}}\left(k'_{\mathrm{in}},k'_{\mathrm{out}}|k{}_{\mathrm{in}},k_{\mathrm{out}}\right)$
and $P_{\mathrm{out}}\left(k'_{\mathrm{in}},k'_{\mathrm{out}}|k{}_{\mathrm{in}},k_{\mathrm{out}}\right)$,
which measure the probability to reach a vertex of degrees $k'_{\mathrm{in}}\mathrm{\, and\,}k'_{\mathrm{out}}$
from a vertex of degrees $k{}_{\mathrm{in}}\mathrm{\, and\,}k{}_{\mathrm{out}}$
following, respectively, an in-link and an out-link. For the particular
case of networks with uncorrelated in-degre and out-degree, $P\left(k_{\mathrm{in}},k_{\mathrm{out}}\right)=P_{\mathrm{in}}(k_{\mathrm{in}})P_{\mathrm{out}}(k_{\mathrm{out}})$
and with uncorrelated degree vertices, such that $P_{\mathrm{i\mathrm{n}}}\left(k'_{\mathrm{in}},k'_{\mathrm{out}}|k{}_{\mathrm{in}},k_{\mathrm{out}}\right)$
and $P_{\mathrm{out}}\left(k'_{\mathrm{in}},k'_{\mathrm{out}}|k{}_{\mathrm{in}},k_{\mathrm{out}}\right)$
are independent of $k{}_{\mathrm{i\mathrm{n}}}\mathrm{\, and\,}k{}_{\mathrm{out}}$
, it can be shown, from a detailed balance relation \cite{Serrano2009},
that:

\begin{equation}
P_{\mathrm{out}}\left(k'_{\mathrm{out}}|k_{\mathrm{out}}\right)=P_{\mathrm{out}}(k'_{\mathrm{out}}).\label{eq: out_degree_degree_correlations}
\end{equation}

The random networks considered in this work are uncorrelated in the
sense defined above.

\section{Mean Field Theory}

\label{3}

We derived an heterogeneous single-site and pair mean-field theory
for the model on a directed network characterized by an out-link degree-degree
correlation $P_{\mathrm{out}}\left(k'_{\mathrm{out}}|k_{\mathrm{out}}\right)$.
Our approach can be easily applied to other network dynamic models.
The single-site approximation time evolution equation for the probability
$p_{k}(C,t)=1-p_{k}(D,t)$ to have a cooperator at a vertex of an
out-degree $k$, at time $t$, is:

\begin{equation}
\frac{d}{dt}p_{k}(C,t)=R_{1}\,\left(1-p_{k}(C,t)\right)-R_{2}\, p_{k}(C,t).\label{eq:single site approx}
\end{equation}
The processes that contribute to the time evolution equation are listed
in Table \ref{table: single site approximation}. The summations in
the rates, $R_{i}$ are over the possible out-degree values of the
neighboring vertices of a given vertex and the average of the imitation
probability over the possible neighborhoods of the vertex and of the
neighboring vertices are taken from binomial distributions, $B_{m}(x)$
where $m$ is the number of attempts (number of neighbors taken into
the average) and $x$ is the success probability of each attempt obtained
from the probability to find a $C$ at a neighboring vertex of a vertex
of a given degree, $k$, at time $t$, $\rho_{k}^{(n)}(t)$ that can
be written as, $\rho_{k}^{(n)}(t)={\textstyle \sum_{m}p_{m}(C,t)P_{\mathrm{out}}(m|k)}$.
Note that for a degree-degree uncorrelated network this quantity does
not depend on the out-degree $k$.

\begin{table}
\centering%
\begin{tabular}{|c|c|c|c|c|}
\hline 
\emph{i} & \emph{process} & \emph{rate,} $R_{i}$ & \emph{n-dist} & \emph{l-dist}\tabularnewline
\hline 
\hline 
$_{1}$ & $_{\cdots\leftarrow D_{k}\rightarrow C_{s}\rightarrow\cdots}$ & $_{\sum_{s}p_{s}(C,t)P_{\mathtt{out}}(s|k)\,\left\langle p_{\mathtt{imitate}}\left(b\left(n+1\right),l+1\right)\right\rangle }$ & $_{B_{k-1}\left(\rho_{k}^{(n)}(t)\right)}$ & $_{B_{s}\left(\rho_{s}^{(n)}(t)\right)}$\tabularnewline
\hline 
$_{2}$ & $_{\cdots\leftarrow C_{k}\rightarrow D_{s}\rightarrow\cdots}$ & $_{\sum_{s}\left(1-p_{s}(C,t)\right)P_{\mathtt{out}}(s|k)\,\left\langle p_{\mathtt{imitate}}\left(n+1,b\, l\right)\right\rangle }$ & $_{B_{k-1}\left(\rho_{k}^{(n)}(t)\right)}$ & $_{B_{s}\left(\rho_{s}^{(n)}(t)\right)}$\tabularnewline
\hline 
\end{tabular}

\caption{Processes and corresponding rates for the single site mean-field approximation
for the model with self-interaction. The quantities $B_{m}\left(x\right)$
represent the Binomial distribution with $m$ attempts and success
probability, $x$.}
\label{table: single site approximation}
\end{table}

The processes that contribute to the pair approximation are listed
in Table \ref{table:pair approximation}. The time evolution equations
for the probabilities, $p_{k,m}(XY,t)$ for a vertex with out-degree
$k$ to follow strategy $X=C,D$ and for a neighbor of that vertex,
with out-degree $m$, to follow strategy $Y=C,D$, at time $t$, are
given by:

\begin{equation}
\begin{array}{lll}
\frac{d}{dt}p_{k,m}\left(CC,t\right) & = & \left(R_{1}+R_{2}\right)\, p_{k,m}\left(DC,t\right)+R_{3}\, p_{k,m}\left(CD,t\right)\\
 &  & \qquad\qquad-\left(R_{4}+R_{5}\right)\, p_{k,m}\left(CC,t\right)\\
\frac{d}{dt}p_{k,m}\left(CD,t\right) & = & R_{10}\, p_{k,m}\left(DD,t\right)+R_{5}\, p_{k,m}\left(CC,t\right)\\
 &  & \qquad\qquad-\left(R_{3}+R_{8}+R_{9}\right)\, p_{k,m}\left(CD,t\right)\\
\frac{d}{dt}p_{k,m}\left(DC,t\right) & = & R_{4}\, p_{k,m}\left(CC,t\right)+R_{6}\, p_{k,m}\left(DD,t\right)\\
 &  & \qquad\qquad-\left(R_{1}+R_{2}+R_{7}\right)\, p_{k,m}\left(DC,t\right)\\
\frac{d}{dt}p_{k,m}\left(DD,t\right) & = & \left(R_{8}+R_{9}\right)\, p_{k,m}\left(CD,t\right)+R_{7}\, p_{k,m}\left(DC,t\right)\\
 &  & \qquad\qquad-\left(R_{10}+R_{6}\right)\, p_{k,m}(DD,t)
\end{array}\label{eq:pair approximation}
\end{equation}

The averages of the imitation probabilities, included in Table \ref{table:pair approximation},
are taken from binomial distributions, as in the single-site approximation
case, with a success probability given by the conditional probability,
$\rho_{k}^{(n)}(X,t)$ to find a $C$ at a vertex neighbor to a vertex
of out-degree $k$ given that the agent at this vertex is following
strategy $X=C,D$. This quantity can be written as:

\begin{equation}
\rho_{k}^{(n)}(X,t)={\textstyle \sum_{m}\frac{p_{k,m}(XC,t)}{p_{k,m}^{(1)}(X,t)}P_{\mathrm{out}}\left(m|k\right)}\label{eq: neighbor conditional prob.}
\end{equation}
where $p_{k,m}^{(1)}(X,t)=p_{k,m}(XC,t)+p_{k,m}(XD,t)$ .

The density of cooperators, $\rho(t)$, is obtained, in the pair approximation,
from:

\[
\rho(t)=\sum_{k,m}p_{k,m}^{(1)}(C,t)P_{\mathrm{out}}(m|k)\, P_{\mathrm{out}}(k)
\]
In the derivation of the pair approximation, probabilities of configurations
of more than two out-linked vertices are approximated by probabilities
of configurations of pairs of out-linked vertices in the spirit of
the probability approximation methods previously applied to dynamical
models on regular lattices \cite{Avraham1992,Ferreira1993,Szabo1998,Szabo2007}
. 

\begin{table}
\centering%
\begin{tabular}{|c|c|c|c|c|}
\hline 
\emph{i} & \emph{process} & \emph{rate,} $R_{i}$ & \emph{n-dist} & \emph{l-dist}\tabularnewline
\hline 
\hline 
$_{1}$ & $_{\cdots\leftarrow D_{k}\rightarrow C_{m}\rightarrow\cdots}$ & $_{\frac{1}{k}\left\langle p_{\mathtt{imitate}}\left(b\left(n+1\right),l+1\right)\right\rangle }$ & $_{B_{k-1}\left(\rho_{k}^{(n)}(D,t)\right)}$ & $_{B_{m}\left(\rho_{m}^{(n)}(C,t)\right)}$\tabularnewline
\hline 
$_{2}$ & $\cdots\leftarrow_{\begin{array}{c}
C_{s}\nwarrow\\
\ldots\swarrow
\end{array}D_{k}\rightarrow C_{m}}$ & $_{\frac{k-1}{k}\sum_{s}\frac{p_{k,s}(DC,t)}{p_{k,s}^{(1)}(D,t)}P_{\mathtt{out}}(s|k)\,\left\langle p_{\mathtt{imitate}}\left(b\left(n+2\right),l+1\right)\right\rangle }$ & $_{B_{k-2}\left(\rho_{k}^{(n)}(D,t)\right)}$ & $_{B_{s}\left(\rho_{s}^{(n)}(C,t)\right)}$\tabularnewline
\hline 
$_{3}$ & $_{C_{k}\rightarrow D_{m}\begin{array}{l}
\nearrow C_{s}\rightarrow\cdots\\
\searrow\ldots
\end{array}}$ & $_{\sum_{s}\frac{p_{m,s}(DC,t)}{p_{m,s}^{(1)}(D,t)}P_{\mathtt{out}}(s|m)\,\left\langle p_{\mathtt{imitate}}\left(b\left(n+1\right),l+1\right)\right\rangle }$ & $_{B_{m-1}\left(\rho_{m}^{(n)}(D,t)\right)}$ & $_{B_{s}\left(\rho_{s}^{(n)}(C,t)\right)}$\tabularnewline
\hline 
$_{4}$ & $\cdots\leftarrow_{\begin{array}{c}
D_{s}\nwarrow\\
\ldots\swarrow
\end{array}C_{k}\rightarrow C_{m}}$ & $_{\frac{k-1}{k}\sum_{s}\frac{p_{k,s}(CD,t)}{p_{k,s}^{(1)}(C,t)}P_{\mathtt{out}}(s|k)\,\left\langle p_{\mathtt{imitate}}\left(n+2,b\, l\right)\right\rangle }$ & $_{B_{k-2}\left(\rho_{k}^{(n)}(C,t)\right)}$ & $_{B_{s}\left(\rho_{s}^{(n)}(D,t)\right)}$\tabularnewline
\hline 
$_{5}$ & $_{C_{k}\rightarrow C_{m}\begin{array}{l}
\nearrow D_{s}\rightarrow\cdots\\
\searrow\ldots
\end{array}}$ & $_{\sum_{s}\frac{p_{m,s}(CD,t)}{p_{m,s}^{(1)}(C,t)}P_{\mathtt{out}}(s|m)\,\left\langle p_{\mathtt{imitate}}\left(n+1,b\, l\right)\right\rangle }$ & $_{B_{m-1}\left(\rho_{m}^{(n)}(C,t)\right)}$ & $_{B_{s}\left(\rho_{s}^{(n)}(D,t)\right)}$\tabularnewline
\hline 
$_{6}$ & $_{D_{k}\rightarrow D_{m}\begin{array}{l}
\nearrow C_{s}\rightarrow\cdots\\
\searrow\ldots
\end{array}}$ & $_{R_{3}}$ &  & \tabularnewline
\hline 
$_{7}$ & $_{D_{k}\rightarrow C_{m}\begin{array}{l}
\nearrow D_{s}\rightarrow\cdots\\
\searrow\ldots
\end{array}}$ & $_{R_{5}}$ &  & \tabularnewline
\hline 
$_{8}$ & $_{\cdots\leftarrow C_{k}\rightarrow D_{m}\rightarrow\cdots}$ & $_{\frac{1}{k}\left\langle p_{\mathtt{imitate}}\left(n+1,b\, l\right)\right\rangle }$ & $_{B_{k-1}\left(\rho_{k}^{(n)}(C,t)\right)}$ & $_{B_{m}\left(\rho_{m}^{(n)}(D,t)\right)}$\tabularnewline
\hline 
$_{9}$ & $\cdots\leftarrow_{\begin{array}{c}
D_{s}\nwarrow\\
\ldots\swarrow
\end{array}C_{k}\rightarrow D_{m}}$ & $_{\frac{k-1}{k}\sum_{s}\frac{p_{k,s}(CD,t)}{p_{k,s}^{(1)}(C,t)}P_{\mathtt{out}}(s|k)\,\left\langle p_{\mathtt{imitate}}\left(n+1,b\, l\right)\right\rangle }$ & $_{B_{k-2}\left(\rho_{k}^{(n)}(C,t)\right)}$ & $_{B_{s}\left(\rho_{s}^{(n)}(D,t)\right)}$\tabularnewline
\hline 
$_{10}$ & $\cdots\leftarrow_{\begin{array}{c}
C_{s}\nwarrow\\
\ldots\swarrow
\end{array}D_{k}\rightarrow D_{m}}$ & $_{\frac{k-1}{k}\sum_{s}\frac{p_{k,s}(DC,t)}{p_{k,s}^{(1)}(D,t)}P_{\mathtt{out}}(s|k)\,\left\langle p_{\mathtt{imitate}}\left(b\left(n+1\right),l+1\right)\right\rangle }$ & $_{B_{k-2}\left(\rho_{k}^{(n)}(D,t)\right)}$ & $_{B_{s}\left(\rho_{s}^{(n)}(C,t)\right)}$\tabularnewline
\hline 
\end{tabular}

\caption{Processes and corresponding rates for the pair mean-field approximation
for the model with self-interaction. The quantities $B_{m}\left(x\right)$
represent the Binomial distribution as in Table \ref{table: single site approximation}.}
\label{table:pair approximation}
\end{table}

\section{out-homogeneous directed network }

\label{4}

We studied the phase diagram of the prisoner's dilemma model in the
out-homogeneous networks using the mean-field approximations and Monte
Carlo simulations. The model exhibits, generically, at lower temptation
$b_{c,1}$, a phase transition from a full cooperation phase to an
intermediate active phase, where both strategies survive, and, at
higher temptation, $b_{c,2}$, a phase transition to a full defection
state. The comparison of the mean-field results with simulations shows
that the single-site approximation is not able to describe correctly
the model and that the pair approximation gives a much better agreement
with simulation.

\begin{figure}
\centering \includegraphics[scale=0.7]{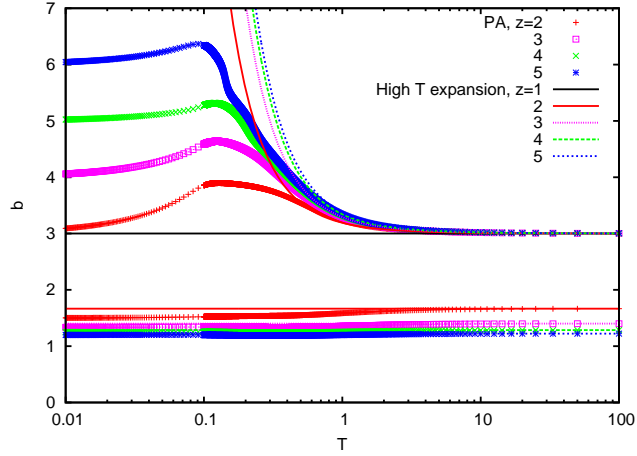} \caption{Phase diagrams in the pair approximation (PA) for the out-homogeneous
networks with $z=2,3,4$ and $5$. The lower curves are $b_{c,1}$
and the upper curves are $b_{c,2}$. The lines are the high temperature
expansion results presented in Eq. (\ref{eq: high T pair approx}). }

\label{fig. phase diagram} 
\end{figure}

\subsection{mean-field approximations }

From a high temperature expansion we were able to obtain analytical
results for the transition parameters, $b_{c,1}$and $b_{c,2}$ for
a network with an arbitrary number of outgoing neighbors, $z$. For
the single-site approximation we obtained $b_{c,1}=\tfrac{2z+1}{2z}$
and $b_{c,2}=2$, and for the pair approximation:

\begin{equation}
\begin{array}{cl}
b_{c,1} & =\frac{2z+1}{2z-1}\\
b_{c,2} & =3+\frac{1}{T}\,\frac{\left(z-1\right)^{2}}{\left(2z-1\right)^{2}}
\end{array}.\label{eq: high T pair approx}
\end{equation}

We also applied the method proposed in Ref. \cite{pedro2015} to numerically
study the MF phase diagram in the temptation and temperature parameters
space. The rate of the time evolution of the density of cooperators
in the equation $\frac{d\rho}{dt}=F(\rho)$ was obtained by solving,
for each $\rho$, a set of modified dynamical equations that conserve
$\rho$: 
\begin{equation}
\begin{array}{ccc}
\frac{d}{dt}p\left(CC,t\right) & = & R(DD,t)\\
\frac{d}{dt}p\left(CD,t\right) & = & -R(DD,t)\\
\frac{d}{dt}p\left(DC,t\right) & = & -R(DD,t)\\
\frac{d}{dt}p\left(DD,t\right) & = & R(DD,t)
\end{array},\label{eq:Modified Dynamics}
\end{equation}
where $R(DD,t)$ is the total rate in the equation for $\frac{d}{dt}p(DD,t)$
in the pair approximation equations (\ref{eq:pair approximation})
and $F(\rho)$ is obtained from the stationary value of $R(CD,t)$
or $R(CC,t)$ , by using $F(\rho)=-R(CD,\infty)=\frac{1}{2}R(CC,\infty)$.

\begin{figure}
\centering\includegraphics[scale=0.7]{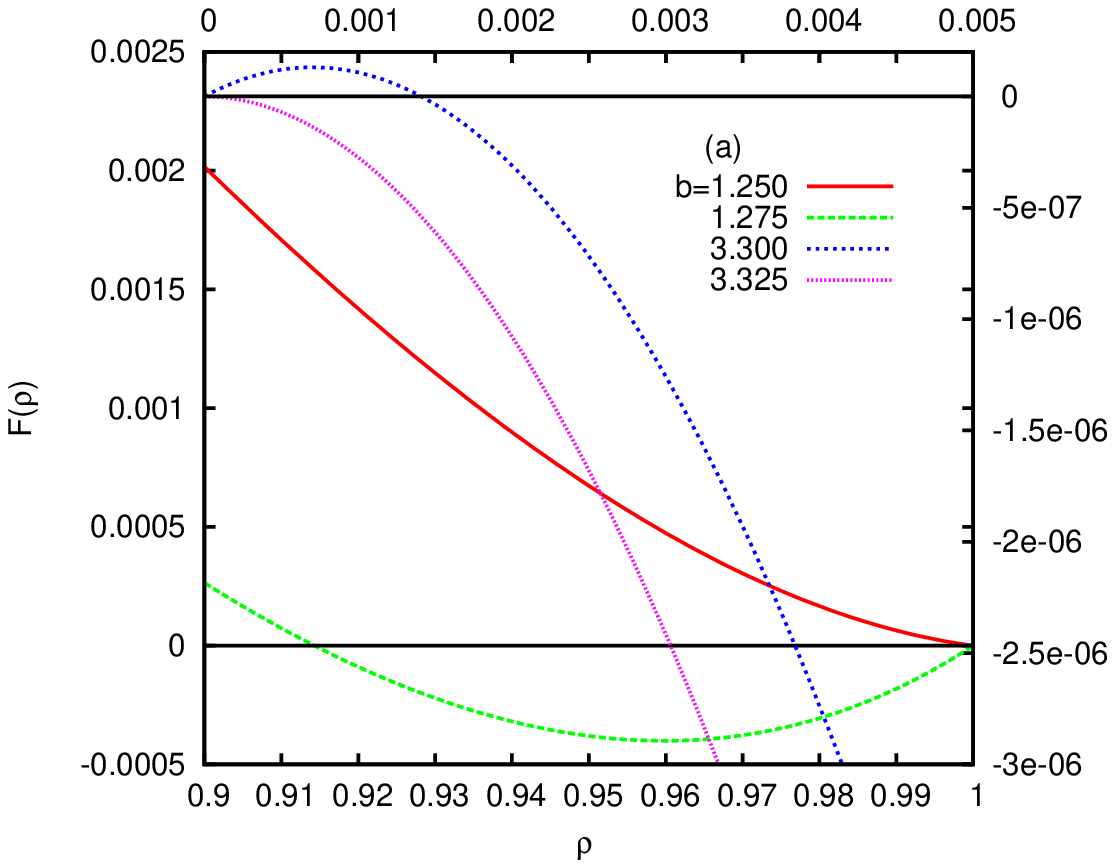}

\centering\includegraphics[scale=0.7]{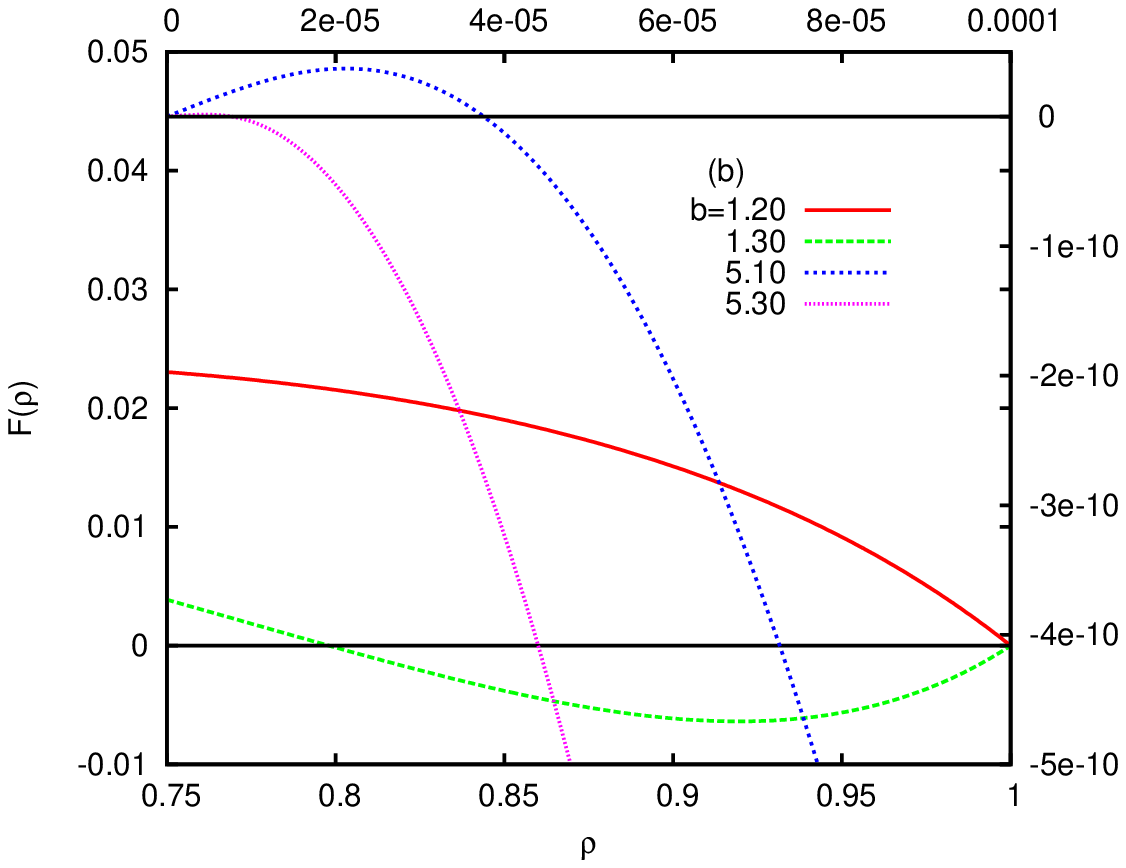}

\caption{Function $F(\rho)$ near the phase transitions at $b_{c,1}$(with
$\rho\sim1$, left and bottom axis) and at $b_{c,2}$ (with $\rho\sim0$,
right and upper axis) for $z=4$ and two temperatures (a) $T=1$ and
(b) $T=0.1$.}
\label{fig: F(rho)}
\end{figure}
The zeros of $\frac{d}{d\rho}F(\rho)$ near $\rho=1$ and near $\rho=0$
give the location of the phase transitions at $b_{c,1}$ and $b_{c,2}$
, respectively. The nature of the phase transition (continuous or
discontinuous) can be obtained from the sign of the second order derivative
of $F(\rho)$ at the phase transition\cite{pedro2015}. The results
obtained by integrating the equations (\ref{eq:Modified Dynamics})
until $\left|R(DD,t)\right|<10^{-12}$, for $z=2,3,4$ and $5$ are
presented in Fig. \ref{fig. phase diagram}. The behavior of $F(\rho)$
near $\rho=1$ and $\rho=0$ with density values separated by $d\rho=10^{-5}$
was used to locate the phase transitions with an error in the temptation
parameter given by $db=10^{-4}$. Both phase transitions are predicted
to be continuous at any temperature and for any coordination number,
$z$ . In agreement with the high temperature expansion results shown
in Eq. (\ref{eq: high T pair approx}) we obtained values of $b_{c,1}$
almost independent of temperature and values of $b_{c,2}$ that increase
for low temperatures from high temperature values close to 3. At low
temperatures the values of $b_{c,2}$ are close to $z+1$. In Fig.
\ref{fig: F(rho)} we show the behavior of $F(\rho)$ near the two
phase transitions for $z=4$ and two temperatures $T=1$ (Fig. \ref{fig: F(rho)}(a)
)and $T=0.1$ (Fig. \ref{fig: F(rho)}(b) ). Note that, at low T,
the determination of the zeros of $F(\rho)$, corresponding to the
transition at $b_{c,2}\sim z+1$, require a high numerical precision
due to the very small value taken by $F(\rho)$ near the origin. From
the lower panel of Fig. \ref{fig: F(rho)} it is clear that the phase
with zero density of cooperators becomes stable only for high values
of the temptation parameter close to $b\sim5.2$.

\subsection{$z=1$ case }

The time integration of the pair mean-field equations and simulations,
for $z=1$, show that the two critical temptations are equal, $b_{c,1}=b_{c,2}=3$,
at any temperature, also in agreement with the high temperature expansion
result Eq. (\ref{eq: high T pair approx}).

\begin{table}
\centering%
\begin{tabular}{|c|c|c|}
\hline 
$i$ & \emph{Configuration} & $\Delta P$\tabularnewline
\hline 
\hline 
$_{1}$ & $_{\cdots\rightarrow D\rightarrow C\rightarrow C\rightarrow\cdots}$ & $_{2-b}$\tabularnewline
\hline 
$_{2}$ & $_{\cdots\rightarrow D\rightarrow C\rightarrow D\rightarrow\cdots}$ & $_{1-b}$\tabularnewline
\hline 
$_{3}$ & $_{\cdots\rightarrow C\rightarrow D\rightarrow D\rightarrow\cdots}$ & $_{-1}$\tabularnewline
\hline 
$_{4}$ & $_{\cdots\rightarrow C\rightarrow D\rightarrow C\rightarrow\cdots}$ & $_{b-1}$\tabularnewline
\hline 
\end{tabular}

\caption{Payoff diferences, $\Delta P=P_{i_{\mu}}-P_{i}$ between strategies
at $\cdots\rightarrow D\rightarrow C\rightarrow\cdots$ (configurations
$1$ and $2$) and $\cdots\rightarrow C\rightarrow D\rightarrow\cdots$
(configurations $3$ and $4$) interfaces for the model with self-interaction
with $z=1$.}
\label{table: z1 Configurations}
\end{table}
This behavior can be understood by looking at the payoff differences,
$\Delta P=P_{i_{\mu}}-P_{i}$ , listed in Table \ref{table: z1 Configurations}
for the possible configurations at the interfaces $D\rightarrow C$
(configurations $1$ and $2$ in Table \ref{table: z1 Configurations})
and $C\rightarrow D$ (configurations $3$ and $4$ in Table \ref{table: z1 Configurations}).
The behavior of the system is controlled by the rates associated with
the motion of the interfaces $\cdots\rightarrow D\rightarrow C\rightarrow C\rightarrow\cdots$
and $\cdots\rightarrow C\rightarrow D\rightarrow D\rightarrow\cdots$
(configurations $1$ and $3$, respectively) because the motion of
the other interfaces lead to the generation of configurations of these
two types. For $b\leq1$ the rates for the displacement of $D\rightarrow C$
interfaces are higher than the rates for $C\rightarrow D$ interfaces.
For $b\geq3/2$, the rate of increase of the number of defectors associated
with configuration $\cdots\rightarrow C\rightarrow D\rightarrow C\rightarrow\cdots$
(configuration $4$) becomes the higher rate in the system but it
generates configurations $\cdots\rightarrow C\rightarrow D\rightarrow D\rightarrow\cdots$
( configuration $3$) which moves, until $b=3$, at a lower rate than
the rate for the spreading of cooperators $\cdots\rightarrow D\rightarrow C\rightarrow C\rightarrow\cdots$
(configuration $1$). Consequently, we expect, for $b<b_{c}=3$, the
system to reach full cooperation and, for $b>b_{c}=3$, the system
to reach full defection. This scenario remains valid at any nonzero
temperature in spite of the temperature dependence of the rates of
motion of the interfaces. This behavior is also in agreement with
the results obtained for the pair approximation and simulations for
$T=10$ and $T=1$ shown in Figs. \ref{fig: Pair Approx and MC beta=00003D0.1 and beta=00003D1}(a)
and (b), respectively. The time integration of the pair approximation
equations reaches full cooperation and full defection very slowly
especially at high temperatures (see the two curves shown in Fig.
\ref{fig: Pair Approx and MC beta=00003D0.1 and beta=00003D1}(a)
obtained for a maximum integration time $10^{6}$ and $10^{7}$ where
full cooperation for $b<b_{c}=3$ and full defection for $b>b_{c}=3$
are still far from being reached). In the same figure we also show
the results of simulations for systems of size $N=10^{3}$ and $N=10^{4}$
showing a strong size dependence but approaching the expected behavior
in the thermodynamic limit. 

\begin{figure}
\centering\includegraphics[scale=0.7]{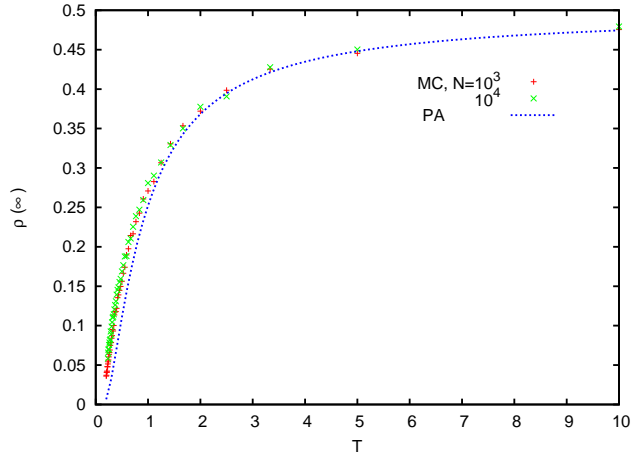}\caption{The stationary density of cooperators, $\rho(\infty)$ for $z=1$
and $b=3$ as a function of temperature from the pair approximation
(PA, Eq. (\ref{eq: z=00003D1 b=00003Dbc stationary rho}) ) and simulations
for systems of size $N=10^{3}$and $N=10^{4}$, taking averages over
$10^{4}$ samples and starting from an equal density of cooperators
and defectors. }
\label{fig: z=00003D1 b=00003D3 Tdep}
\end{figure}

For the pair approximation, we have found that, exactly at $b=b_{c}=3$,
$\rho(\infty)$ depends on temperature and initial condition and is
given by the following expression:

\begin{equation}
\rho(\infty)=\frac{\left({e}^{2\,\beta}\lyxmathsym{\textminus}{e}^{\beta}\right)\,{\rho\left(0\right)}^{2}+\left({e}^{\beta}+1\right)\,\rho\left(0\right)}{{e}^{2\,\beta}\text{+}1-\left({e}^{3\,\beta}\lyxmathsym{\textminus}{e}^{2\,\beta}+{e}^{\beta}\lyxmathsym{\textminus}1\right)\,\rho\left(0\right)\left(\rho\left(0\right)-1\right)}.\label{eq: z=00003D1 b=00003Dbc stationary rho}
\end{equation}
Results from simulations presented in Fig. \ref{fig: z=00003D1 b=00003D3 Tdep}
agree with the pair approximation result (Eq (\ref{eq: z=00003D1 b=00003Dbc stationary rho}))
at high temperatures and show some deviation at low temperatures. 

At $T=\infty$, each agent adopts the strategy of a randomly selected
neighbor with probability $\frac{1}{2}$ and the model is related
to the direct voter model on directed networks considered in Ref.
\cite{Serrano2009}. In this limit, the density of cooperators is
conserved and it remains equal to its initial value. 

At $T=0$ the interface dynamics becomes deterministic and some of
the interfaces are frozen while others are moving depending on the
value of $b$ (only interfaces with $\Delta P\geq0$ move) . For $b\leq1$
only the $D\rightarrow C$ interfaces are active and the system should
reach full cooperation (except for frozen initial domains of defectors
with a negligible weight in large systems). For $b>1$ the interface
$\cdots\rightarrow D\rightarrow C\rightarrow D\rightarrow\cdots$
is frozen and defectors with a neighbor cooperator that has a neighbor
defector start to survive and for $b\geq2$ the only active interface
corresponds to the configuration $\cdots\rightarrow C\rightarrow D\rightarrow C\rightarrow\cdots$.
After the disappearance of all cooperators with a neighbor defector
that has a neighbor cooperator the system will reach a frozen disordered
state, free of such configurations,  with a final density of cooperators
that depends on its value on the starting configuration.

\subsection{$z>1$ case }

\begin{figure}
\centering\includegraphics[scale=0.7]{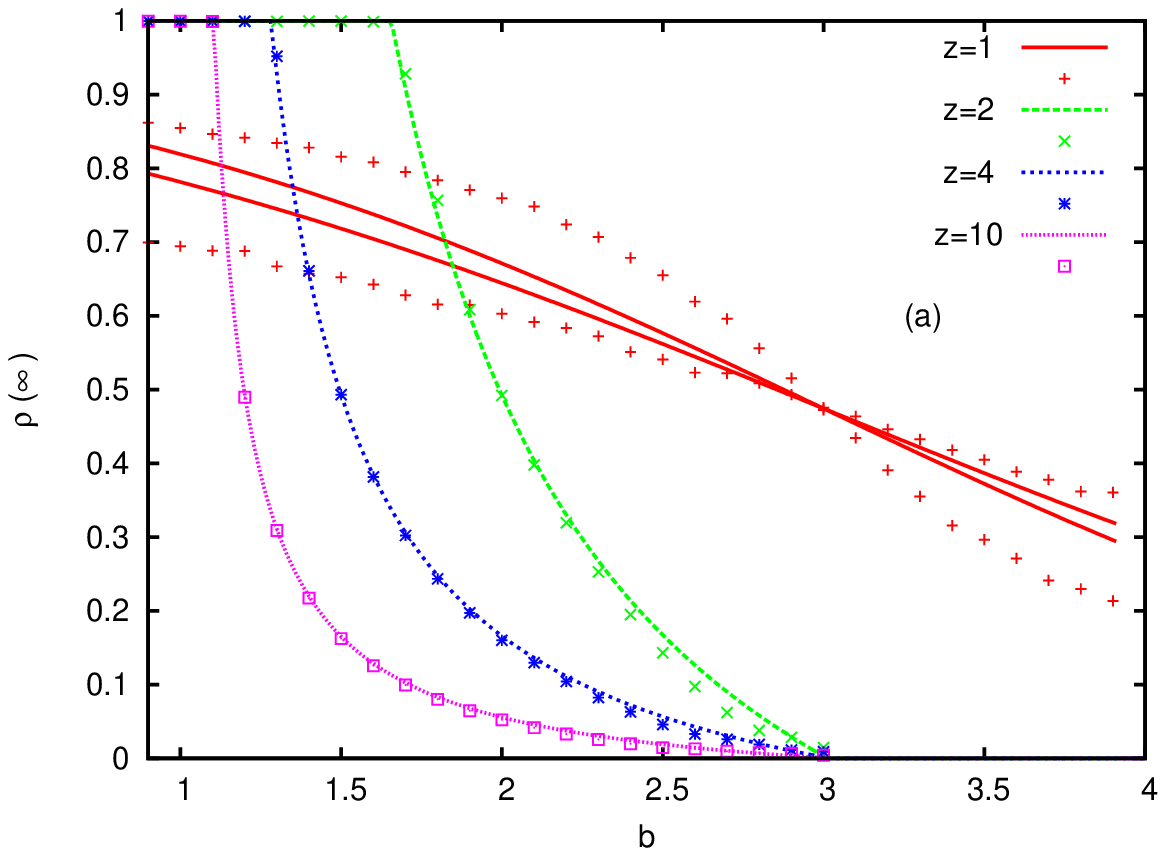}

\centering\includegraphics[scale=0.7]{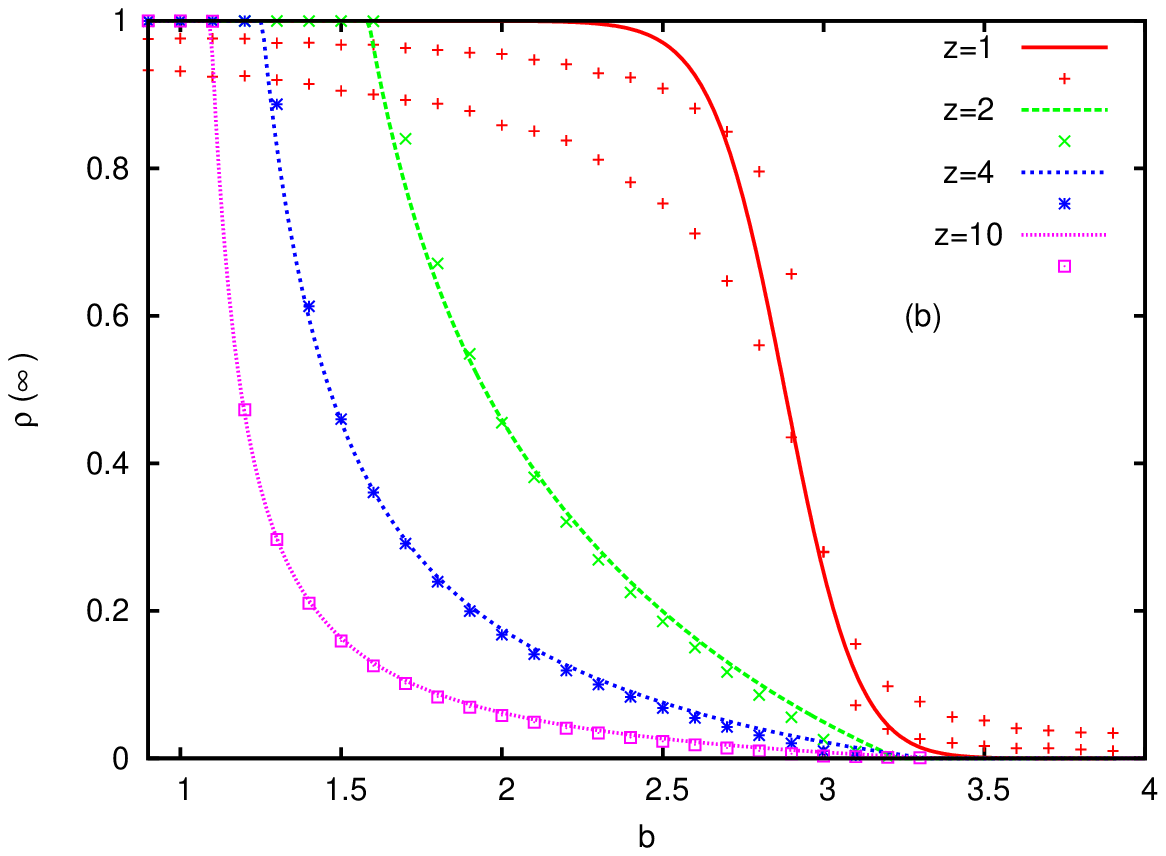}

\caption{Comparison of the predictions of the pair approximation (lines) and
Monte Carlo simulations (points) for the stationary density of cooperators
$\rho(\infty)$ for (a) $T=10$ and (b) $T=1$. The simulations were
done for systems of size $N=10^{4}$ and averages were made over $100$
samples. For $z=1$ simulations are shown for system sizes $N=10^{3}$and
$10^{4}$ taking averages over $10^{4}$ samples. For the pair approximation
the $z=1$ results are not stationary and correspond to cooperator
densities reached at integration times $10^{6}$ and $10^{7}$in (a)
and $10^{6}$in (b).}
\label{fig: Pair Approx and MC beta=00003D0.1 and beta=00003D1}
\end{figure}

For networks with a number of outgoing links, $z$, larger than one
there is a region of coexistence of strategies for $b$ values between
the two critical parameters $b_{c,1}$ and $b_{c,2}$. In Fig. \ref{fig: Pair Approx and MC beta=00003D0.1 and beta=00003D1}
we show, for $z=2,\,4$ and $10$ the stationary density of cooperators
obtained by simulations of systems of size $N=10^{4}$ for $T=10$
(Fig. \ref{fig: Pair Approx and MC beta=00003D0.1 and beta=00003D1}(a)
) and $T=1$ (Fig. \ref{fig: Pair Approx and MC beta=00003D0.1 and beta=00003D1}(b)
) taking averages over $100$ samples. In all our simulations the
stationary averages were obtained by neglecting the initial transient
time dependence and averaging only over samples that have not reached
any of the two absorbing states (full cooperation and full defection).
We also plot the pair approximation predictions showing a good agreement
with MC simulations except near the phase transitions where slightly
different values for the critical temptation parameters are obtained.
In the pair approximation, for the network with $z=4$ at $T=1$ we
got $b_{c,1}=1.2542(1)$ and $b_{c,2}=3.3241(1)$ and results from
simulations give $b_{c,1}=1.274(1)$ and $b_{c,2}=3.14(1)$. Furthermore,
the agreement between simulations and the pair approximation worsens
as the temperature decreases.

The payoff differences at $D\rightarrow C$ interfaces take the possible
values $\Delta P_{D\rightarrow C}=m+1-n\, b$ with $0\leq m\leq z$
and $1\leq n\leq z$ and at $C\rightarrow D$ interfaces take the
values $\Delta P_{C\rightarrow D}=n\, b-m-1$ with $0\leq m\leq z-1$
and $0\leq n\leq z$, where $m$ and $n$ are the possible number
of cooperators in the neighborhood of the cooperator and of the defector,
respectively. At low temperatures, the dynamics at interfaces with
negative payoff differences is slow ( freezing at $T=0$) which generates
plateaus in the density of cooperators in between the values of temptation,
$b_{n,m}=\frac{m+1}{n}$, where the payoffs for the interfaces change
sign. 

\begin{figure}
\centering\includegraphics[scale=0.7]{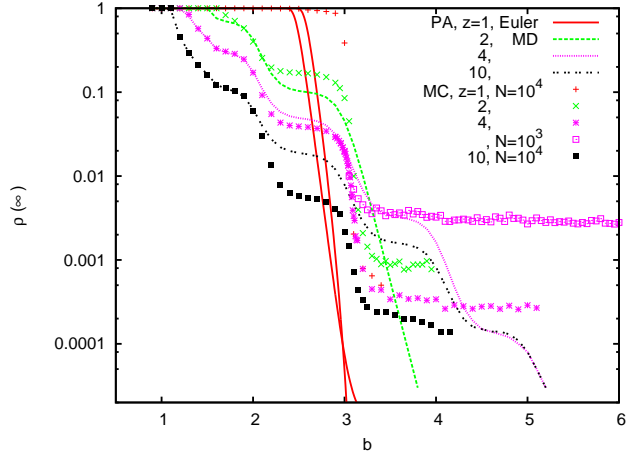}

\caption{We compare the temptation dependence of the stationary density of
cooperators, $\rho(\infty)$ , for pair approximation (PA) and Monte
Carlo simulations (MC) at a low temperature, $T=0.1$, for networks
with $z=1,\:2,\:4$ and $10$. The PA results for $z=1$ were obtained
from time integration for $10^{6}$and $10^{7}$ steps showing a slow
approach to the stationary behavior. Results for other values of $z$
were obtained from Eq. \ref{eq:Modified Dynamics}(MD) . The MC results
were obtained for systems of size $N=10^{4}$ and $N=10^{3}$ (for
$z=4$) to show the finite size effects at large values of $b$, for
very small densities of cooperators.}

\label{fig: pair approximation and MC beta=00003D10}
\end{figure}

Near the full cooperator state the most important configurations for
the $D\rightarrow C$ interfaces correspond to $m=z$ and $n=z$ which
becomes inactive ( stopping creating cooperators), at very low T,
for $b>b_{z,z}=1+1/z$ meaning that an isolated $D$ in a sea of $C$
start surviving. The most important configuration for the $C\rightarrow D$
interface, near the full cooperator state, correspond to $m=z-1$
and $n=z$ which becomes active (starting creating defectors), at
very low T, for $b>b_{z,z-1}=1$. These defectors will be at new $C\rightarrow D$
interfaces where the defector now has a neighboring defector and the
new interface is active and leading to the creation of new defectors
for $b>b_{z-1,z-1}=1/(1-1/z)\sim1+1/z$ . This explains the observation
of values of $b_{c,1}$ close to 1, in agreement with the high temperature
expansion, Eq. (\ref{eq: high T pair approx}), which gives a dependence
on the network coordination number approaching 1 in the limit of an
infinite number of neighbors.

The most important configurations for $D\rightarrow C$ interfaces,
near the full defection state, correspond to $m=0$ and $n=1$ with
$\Delta P_{D\rightarrow C}=1-b$ and for the $C\rightarrow D$ interface
$m=0$ and $n=0$ with $\Delta P_{C\rightarrow D}=-1$ . However the
motion of the $D\rightarrow C$ interface with $m=0$ and $n=1$ generates
a new cooperator which will be at a $D\rightarrow C$ interface with
$m=1$ and $n=1$ corresponding to $\Delta P_{D\rightarrow C}=2-b$
. Thus, like in the case for $z=1$, we would expect that the processes
that generate defectors will win over those that generate cooperators
for $b>b_{c,2}\sim3$ . At very low T, since cooperators without any
neighboring cooperators always have higher payoffs than defectors
also without neighboring cooperators, independently of $b$, the interfaces
$C\rightarrow D$ will be sluggish, and cooperators may persist in
the system for long times. Furthermore, clusters of cooperators with
a $C$ connected to other $C$ resist as long as the weakest cooperators,
in the periphery of the cluster, with no neighboring cooperators resist
invasion by defectors. The $C$ in the root of a cluster connected
to $z$ cooperators is a source of new $C$ up to $b=z+1$. The newly
generated cooperators, in the new roots, have only one neighboring
$C$ and, for values of $b>2$, do not generate further spreading
of cooperators. At a finite, low $T$, the balance between the rate
of disappearance of $C$ in the periphery of the cluster, where the
cooperators have no neighboring cooperators, and the rate of creation
of new $C$, from a $C$ with only one neighboring $C$ remains the
most important balance leading to $b_{c,2}=3.$ However, the pair
approximation is not sensitive to the weakness of the newly generated
cooperators and shows incorrectly, at low temperatures, $b_{c,2}\sim z+1$,
which is the temptation limit above which, at zero temperature, all
the processes generating cooperators stop.

\begin{figure}
\centering\includegraphics[scale=0.7]{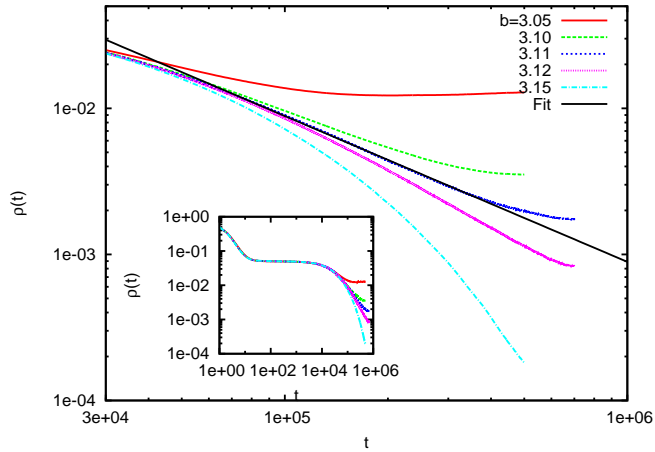}

\caption{Monte Carlo simulations of the time dependence of the density of cooperators,
$\rho(t)$, at a low temperature $T=0.1$ for a network with $z=4$
with a system size $N=10^{6}$ showing that $b_{c,2}=3.11(1)$, much
smaller than the value predicted by the pair approximation. The power
law fit of $\rho(t)$, for large times, shown in the plot, has a slope
$\alpha=-0.998$. In the inset we present the full curve $\rho(t)$
including the behavior at short times.}

\label{fig: Rho_t_Bc2ForT=00003D0.1z=00003D4}
\end{figure}

In Fig. \ref{fig: pair approximation and MC beta=00003D10} we show
for the low temperature, $T=0.1$, the stationary density of cooperators,
$\rho(\infty)$, as a function of $b$, obtained from simulations
and the pair approximation for systems of $z=2,\,4$ and $10$. The
pair approximation results were obtained from the modified dynamics
(MD) Eq. (\ref{eq:Modified Dynamics}) and time integration of the
mean-field equations (Eq. (\ref{eq:pair approximation})) (for $z=1$).
The agreement of the pair approximation with simulations is good for
small values of $b$, but the height of the plateaus in $\rho(\infty)$,
for $b\geq2$, when $\rho(\infty)\lesssim0.1$, are not correctly
predicted by the pair approximation. The simulation results for $z=4$
and sizes $N=10^{3}$ and $10^{4}$ show that for $b\gtrsim3$ the
observed plateaus have heights that decrease with the size of the
system suggesting that in the thermodynamic limit the system reaches
the full defection absorbing state. We determined $b_{c,2}$ from
simulations of the time dependent behavior of $\rho(t)$ for a network
with $z=4$ and size $N=10^{6}$ (see Fig. \ref{fig: Rho_t_Bc2ForT=00003D0.1z=00003D4})
obtaining $b_{c,2}=3.11(1)$ which is much smaller than the value
$b_{c,2}\sim5$ predicted by the pair approximation. For large times
we observe $\rho(t)\sim1/t$ as expected for the mean-field universality
class of phase transitions to a single absorbing state\cite{dickman99}.

It is interesting to comment on the behavior of the system for increasing
values of $z$: for very large $z$, we approach the fully connected
network limit, where it is known that the system jumps from full cooperation
to full defection at $b=1$. Our results show that for large $z$,
the critical temptation $b_{c,1}$ gets closer to $1$ and the density
of cooperators decays very strongly for $b>b_{c.1}$ but still shows
a small nonzero value for temptation values up to $b_{c,2}$ which
does not approach $b_{c,1}$ for arbitrary large $z$.

\subsection{model without self-interaction }

We can repeat the above arguments for the model without self-interaction
and conclude that for $z=1$ we have, $b_{c,1}=b_{c,2}=1$ with a
suppression of cooperation for $b>1$ . For larger $z$ we have now
$b_{n,m}=\frac{m}{n}$ and near the full cooperator state a single
$D$ at a $D\rightarrow C$ interface start surviving, at low $T$,
for $b>b_{z,z}=1$. The $C\rightarrow D$ interfaces with $m=z-1$
and $n=z$ lead to new $C\rightarrow D$ interfaces with $m=z-1$
and $n=z-1$ which are active and leading to the proliferation of
defectors, for $b>b_{z-1,z-1}=1$. Consequently, we expect $b_{c,1}\sim1$.
Considering now configurations close to the full defection state we
see that a single cooperator at a $C\rightarrow D$ interface in a
sea of defectors has $\triangle P_{C\rightarrow D}=0$ and such interface
is always active. The $D\rightarrow C$ interface for the cooperator
in the sea of defectors $(m=0$ and $n=1$ ) has $\triangle P_{D\rightarrow C}=-b$
and it is inactive at very low $T$. The production of $C$ occurs
predominantly at configurations where a $C$ has a neighboring $C$
$(m=1$ and $n=1$ ) with $\triangle P_{D\rightarrow C}=1-b$ and
the balance between $C$ production and $D$ production turns in favor
of defectors for $b>b_{c,2}\sim1$. From an high temperature expansion
(weak selection limit) we obtained from the pair approximation:

\begin{equation}
\begin{array}{cl}
b_{c,1} & =1-\frac{1}{T}\,\frac{\left(z-1\right)^{2}}{8z^{3}-12z^{2}+6z-1}\\
b_{c,2} & =1+\frac{1}{T}\,\frac{\left(z-1\right)^{2}}{\left(2z-1\right)^{2}}
\end{array}.\label{eq: high T pair approximation Model without Self}
\end{equation}

\begin{figure}
\centering\includegraphics[scale=0.7]{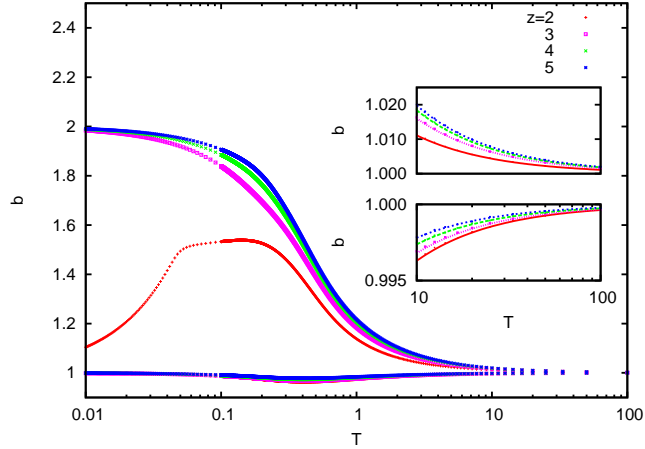}\caption{Phase diagram of the model without self-interaction, in the pair approximation,
for the out-homogeneous networks with $z=2,3,4$ and $5$. The lower
curves are $b_{c,1}$ and the upper curves are $b_{c,2}$, respectively.
The inset shows the high temperature region where the numerical results
based on Eq. (\ref{eq:Modified Dynamics}) are compared with the analytical
results (lines) presented in Eq. (\ref{eq: high T pair approximation Model without Self}).}

\label{fig: pair approx phase diagram without self-interaction}
\end{figure}

Contrary to the case with self-interaction a cluster of $C$ in a
sea of $D$ is not stable at low T because now the cooperators in
the periphery (with no neighboring $C$) do not resist invasion by
defectors. Consider the cluster $D\rightarrow C\rightarrow C\rightarrow D$
with a $D\rightarrow C$ and a $C\rightarrow D$ interface and consider
that the remaining neighbors are all $D$ . At $T=0$, the $D\rightarrow C$
interface is frozen for $b>1$ and the activity at the interface $C\rightarrow D$
leads to the disappearance of the cooperators. For the cluster to
be sustainable there is need for the interfaces $C\rightarrow D$
in the periphery to move at a lower rate than the interfaces $D\rightarrow C$
at the root which leads to the balance being in favor of defectors
for $b>1$. When the defector at the $D\rightarrow C$ interface is
facing a $C$ with, for example, two neighboring $C,$ then cooperators
are generated with higher rate determined by the payoff difference
$2-b$ but still the newly generated $C$ will have only one $C$
neighbor and again for $b>1$ defectors will win. In Fig.\ref{fig: pair approx phase diagram without self-interaction}
we show the pair approximation phase diagram using the same method
previously applied to the model with self-interaction. In the inset
we see that the high temperature expansion Eq. \ref{eq: high T pair approximation Model without Self}
agrees well with the numerical results.

\begin{figure}
\centering\includegraphics[scale=0.7]{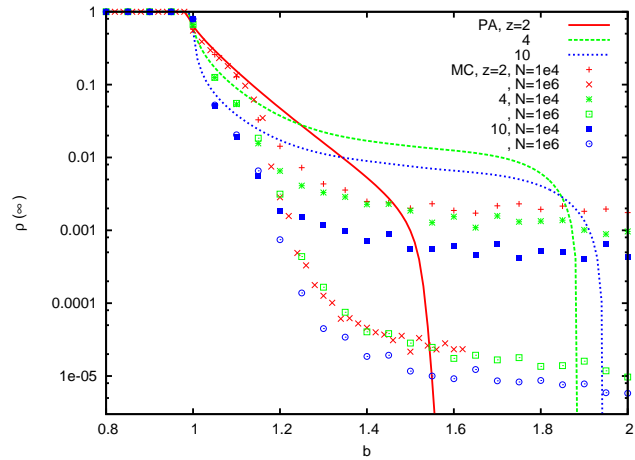}

\caption{Comparison of pair approximation (PA) and MC simulation results for
the steady state density of cooperators for the model without self-interaction
at a low temperature $T=0.1$ and for out-homogeneous directed networks
with $z=2$, $4$ and $10$.}

\label{fig: Low T=00003D0.1 MC and Pair approximation}
\end{figure}

At very low $T$ we obtain for the pair approximation a value $b_{c,2}\sim1$
for $z=2$ and $b_{c,2}\sim2$ for $z>2$. This difference in behavior
may be related with the different relative statistical weight given
by the pair approximation to configurations with a $C$ with one and
two $C$ as neighbors for $z=2$ and $z>2$ because if the configurations
with two neighboring $C$ are predominant we expect the rate of creation
of $C$ to be determined by the payoff difference $2-b$ leading to
$b_{c,2}\sim2$ . We made MC simulations of networks with, $z=2,4$
and $10$, at a low temperature, $T=0.1$ , and the results shown
in Fig. \ref{fig: Low T=00003D0.1 MC and Pair approximation} correspond
to a value $b_{c,2}\sim1.2$ much smaller than the pair approximation
prediction.

\section{Random directed networks }

\label{sec:5} 
\begin{figure}
\centering\includegraphics[scale=0.7]{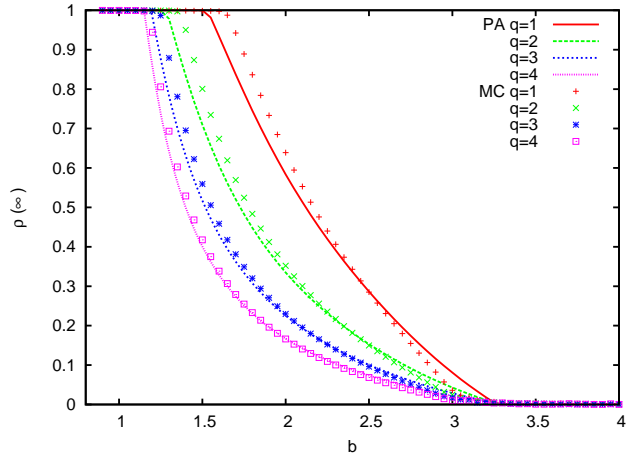}

\caption{Comparison of the pair approximation (PA) and simulations (MC) of
the PD model with self-interaction for uncorrelated directed random
networks with size $N=10^{4}$, with average number of neighbors,
$\left\langle z\right\rangle =1.582$ (q=1), $2.313$ (q=2), $\left\langle z\right\rangle =3.1572$
(q=3) and $\left\langle z\right\rangle =4.0746$ (q=4). }

\label{fig: Erdos-Renyi T=00003D1}

\end{figure}

In random directed networks the distribution of the
number of outgoing links of a given vertex is Poissonian and strongly
peaked at its average number. The behavior of the stationary density
of cooperators is expected to follow the same trend as for out-homogeneous
directed random networks with a given number of outgoing links. To
test the accuracy of the mean-field pair approximation in Eq. (\ref{eq:pair approximation})
for an heterogeneous case we compared it with MC simulations, for
the model with self-interaction at $T=1$ for different average number
of out-neighbors. The results in Fig. \ref{fig: Erdos-Renyi T=00003D1}
show that the pair approximation provides a reasonably good approximation
for the behavior of the system. Since we use a modified Poissonian
distribution, as described in section \ref{2}, the limit $q\rightarrow0$
of these networks corresponds exactly to out-homogeneous networks
with $z=1$.

\section{Concluding Remarks}

\label{sec:6}

We have derived an heterogeneous pair mean-field approximation that
is able to correctly describe the behavior of the prisoner's dilemma
model in directed networks in the limit of high T (weak selection).
At low T the pair approximation gives predictions for the critical
parameter $b_{c,2}$ in disagreement with MC simulation. The pair
approximation was numerically studied by using a very efficient method,
previously applied to other models\cite{pedro2015}, which is based
on modified dynamical equations that solve the time-dependent equations
under the constraint of a given density of cooperators. We also obtained
analytical expressions for the two critical parameters $b_{c,1}$
and $b_{c,2}$ in the limit of high temperature. For the model with
self-interaction $b_{c,1}$ has a weak temperature dependence and
approaches $1$ for large $z$ and $b_{c,2}$ is close but larger
than $3$ for any $z$. Without self-interaction, at any temperature,
$b_{c,1}$ is close but smaller than $1$ and $b_{c,2}$ is greater
but close to 1 . Essentially, the inclusion of the self-interaction
promotes a shift of $b_{c,2}$ from values close to $1$ to values
close to $3$. The case $z=1$ is a special case where, in both cases,
$b_{c,1}=b_{c,2}$, independent of temperature. In the model without
self-interaction we find $b_{c,1}=b_{c,2}=1$ and cooperation is fully
suppressed for $b>1$ while in the model with self-interaction the
full cooperator state is still reached for temptation values up to
$b_{c,1}=b_{c,2}=3$. In both models the coexistence region does not
shrink as $z$ increases but the levels of cooperation strongly decrease
with increasing $z$, as the networks get closer to the fully connected
limit. This behavior is similar to the one previously reported for
one-dimensional regular lattices with varying coordination number\cite{pacheco2005}.
We can also compare our results to other available results for non-directed
lattices and random networks. The inclusion of self-interaction increases
$b_{c,2}$ from $\sim1.035$ to $\sim1.85$, at $T=0.1$ in the square
lattice (see Refs \cite{Szabo1998,Szabo2005}) while in our out-homogeneous
directed networks we observe a much larger increase in $b_{c,2}$.
In previous studies of non-directed lattices and random regular graphs\cite{Szabo2005,Vukov2006}
two types of phase diagrams were found: (1) one where $b_{c,2}$ shows
a non-monotonous dependence with $T$ , being equal to $1$ at $T=0$
and $T=\infty$ and (2) another type where $b_{c,2}$ decreases with
$T$ showing the highest value in the noise-free limit $T=0$ . Our
MC simulation results suggest that the phase diagram in the directed
networks studied here are of the first type showing an equal high
and low $T$ limits for $b_{c,2}$ although not equal to $1$ in the
case of the model with self-interaction. It is not clear if it is
possible to observe phase diagrams of type (2) in directed networks.
Our main results were obtained for out-homogeneous networks but we
expect the main conclusions to apply also to directed random networks
with Poissonian in and out degree distributions. 

The study of evolutionary dynamics in other types of directed networks
and in mixed directed/ non-directed networks taking into consideration
the difference between interaction and learning/reproduction networks
may be relevant for  the modeling of real systems.

\section*{Acknowledgements}

We acknowledge support from the joint bilateral project FCT/1909/27/2/2014/S
and CAPES 385/14. This work was also partially funded by FEDER funds
through the COMPETE 2020 Programme and National Funds throught FCT
- Portuguese Foundation for Science and Technology under the project
UID/CTM/50025/2013. W.F. also acknowledge the support of the Brazilian
agency CNPq, Grant no. 2013/303253-4. The research of A. L. was supported
by Narodowe Centrum Nauki (NCN, Poland) Grant No. 2013/09/B/ST6/02277.

\bibliographystyle{iopart-num}
\addcontentsline{toc}{section}{\refname}\bibliography{references}
 
\end{document}